\documentclass[twocolumn,showpacs,amsmath,amssymb]{revtex4}
\usepackage{amssymb}
\usepackage{txfonts}
\usepackage{bbm}
\usepackage{graphicx}
\usepackage{appendix}
\usepackage{epsf}
\usepackage{epstopdf}

\begin{document}

\title{A minimum single-band model for low-energy excitations in superconducting K$_x$Fe$_2$Se$_2$ }

\author
{Tao Zhou$^{1}$ and Z. D. Wang$^{2}$}

\affiliation{$^{1}$College of Science, Nanjing University of Aeronautics and Astronautics, Nanjing 210016, China\\
$^{2}$Department of Physics and Center of Theoretical and Computational Physics, The University of Hong Kong,
Pokfulam Road, Hong Kong, China}

\date{\today}
\begin{abstract}
We propose a minimum single-band model for the newly discovered iron-based superconducting  K$_x$Fe$_2$Se$_2$. Our model is found to be numerically consistent with the five-orbital model at low energies. Based on our model and the random phase approximation, we study the spin fluctuation and the pairing symmetry of superconducting gap function. The $(\pi/2,\pi/2)$ spin excitation and the $d_{x^2-y^2}$ pairing symmetry are revealed. All of the results can well be understood  in terms of the interplay between the Fermi surface topology and the local spin interaction, providing a sound picture to explain why the superconducting transition temperature is as high as to be comparable to those in pnictides and some cuprates. A common origin of superconductivity is elucidated for this compound and other high-T$_c$ materials.

\end{abstract}
 \maketitle

The iron-pnictide materials have attracted much attention since their discovery~\cite{kam} due to the high superconducting (SC) transition temperature and qualitatively similar phase diagram with the cuprates. The ground state of the parent compound exhibits a collinear antiferromagnetic (AF) order and the superconductivity emerges when the magnetic order is suppressed with the doping. Thus
 the interplay between the magnetism and superconductivity is likely rather essential for exploring the mechanism of the superconductivity in this class of materials [see e.g., Refs.~\cite{sad,ish}]. The magnetism in the parent compound can be understood from an inter-band Fermi surface (FS) nesting picture, namely, the nesting between the hole and electron bands could give rise to the collinear AF magnetic order~\cite{jiang,zho}. On the other hand, the collinear AF order may also be obtained based on a $J_1-J_2$ Heisenberg model with $J_2>J_1/2$~\cite{yil,fma,qsi}. For the doped systems, the superconductivity with $s_\pm$ pairing symmetry was proposed~\cite{mazin}, which is based on a scenario that
the superconductivity is meditated by the magnetic fluctuation and the inter-band FS nesting still plays a key  role~\cite{mazin,yao}.

Recently, a new type of iron-based superconducting compounds K$_x$Fe$_{2}$Se$_2$ was reported, with the maximum SC transition temperature above 30K ~\cite{guo,mazio,fang}. The crystal structure is similar to that of the previous 122-type iron pnictides, with intercalating potassium atoms between
quasi-two-dimensional conducting Fe-Se layers.
%While some unusual properties were revealed.
It is a heavily over-doped compound with the electron doping about $0.5$ with respect to the parent compound. Both the angle-resolved photoemission spectroscopy (ARPES) experiments~\cite{zhang,mou,zhao,xpw,qian} and the local density approximation (LDA) calculations~\cite{she,yan,cao} indicated that the hole FS pockets disappear and only electron FS pockets exist. This seems consistent with previous ARPES experimental results for iron pnictides that the hole-like FS pockets disappear at  the doping $\delta\sim 0.15$~\cite{sek}. For the doping $\delta\sim 0.5$, the hole-like band should be well below the Fermi energy and make no contribution to the superconductivity. Obviously, the inter-band FS nesting picture, which seems to work well for iron pnictides, is unable to straightforwardly account for the superconductivity of this compound, and thus the previous physical picture for iron-based SC materials meets a serious challenge. A second challenge is to answer why the superconductivity and magnetism could survive at so heavily overdoped compound. For this, it is natural and important to ask whether the interplay between magnetism and superconductivity is still essential for the superconductivity of this compound. Moreover, The pairing symmetry is also an important and critical issue to be addressed, which may be different from that of iron pnictides due to the absence of the hole FS pockets. Actually, the ARPES experiments~\cite{zhang,mou,zhao,xpw} indicated that the gap function is fully gapped for the pockets around $M$-point while it is small near the $\Gamma$ point~\cite{zhang,mou}, which is sharp contrast to the $s_{x^2y^2}$-wave symmetry. Up to now, the pairing symmetry is still unclear: both $s$-wave and $d$-wave symmetry have been proposed~\cite{yu,wang,mai,kot,sai,yzhou}.

From the band calculations for this compound~\cite{she,yan,cao},
it seems that all of the five orbitals hybridize strongly, and thus it was proposed that all
five d-orbitals should be considered to construct a model
~\cite{wang,yu}. Note that the five-orbital model is one of the most frequently-used model to describe the iron-pnictide compound~\cite{kur}.
However, this model includes some non-essential parts that may be redundant for superconductivity (especially those bands far away from the Fermi surface), and even
  make it quite difficult to accurately analyze and understand certain essential physics behind because too many unknown parameters are involved. For example,
if the electron correlation needs to be included, both inter-orbital and intra-orbital interactions should be taken into account;
while the corresponding interaction strengths are difficult to be determined (or adjusted) in order to figure out the relevant physics.
Therefore, it is highly-demanded and desirable to establish a rather simplified minimum model that is able to capture the essential features of this compound. Motivated by this, a minimum two-band model was proposed for iron pnictides~\cite{jiang,rag}, and many physical properties have been understood based on this type of two-band model, especially for those related closely to the low energy excitations.
When studying the K$_x$Fe$_2$Se$_2$ materials, a similar two-orbital model was put forward, while the obtained FS size is much larger than that obtained from the band calculation~\cite{ryu}. Intriguingly, as discussed below, we can refine the minimum model of this compound to have one band to capture the essential physics with better results. As seen from the band calculations~\cite{she,yan,cao} there exist two electron-like FS sheets around $M=(\pi,\pi)$ point. Around the $\Gamma=(0,0)$, a small FS pocket may exist, but it is k$_z$ dependent and electron-like. The FSs from ARPES experiments are consistent with the LDA calculations while the electron pockets around the $\Gamma$ point have a very low spectral weight. So it is reasonable to believe that only the FS pockets around M point are essential in the minimum model, while all other bands that do not cross the Fermi energy may be neglected.  Taking into account that one unit cell consists of two iron irons plus the band folding effect, only one kind of FS pocket around the $X$ or its symmetric points is actually relevant in the unfolded BZ.

In this Letter, motivated by the above considerations, we propose a single-band tight-banding model as a minimum one for the superconducting K$_x$Fe$_2$Se$_2$~\cite{note}. We show that our model could have a similar band dispersion crossing the FS with that of the five-orbital model fitted from the LDA band calculations for the K$_x$Fe$_2$Se$_2$ materials~\cite{yu}. Thus the present model serves as an effective one for describing the low energy physics of the K$_x$Fe$_2$Se$_2$. The spin susceptibility is calculated and analyzed based on the FS topology. It is found that the FS topology may connect to the $J_1-J_2-J_3$ couplings to provide a coherent picture for the spin-density-wave (SDW) instability of parent compound and the magnetic fluctuation in the SC state. Our numerical results present a natural explanation for the high SC transition temperature of this material. The SC gap is also calculated self-consistently and the robust $d$-wave SC pairing is revealed. The $d$-wave symmetry can be explained well based on the spin fluctuation picture and the fermiology theory.

We start from a $t-J$ type model including the tight-banding term and the local spin interaction, which reads
\begin{equation}
H=\sum_{{\bf k}\sigma}\varepsilon_{\bf k}n_{{\bf k},\sigma}+H_J,
\end{equation}
where $\varepsilon_{\bf k}$ is taken phenomenologically as the single band tight-banding approximation: $\varepsilon_{\bf k}=-2t(\cos kx+\cos ky)-4t^{\prime}\cos k_x\cos k_y-\mu$ with $t=0.08$ eV and $t^{\prime}=-0.2$ eV. The chemical potential $\mu$ is controlled by the electron doping $\delta$. $H_J$ is the local spin interaction.
From the first principle calculation in Ref~\cite{yan}, we assume the spin interaction
to include the nearest-neighbor, the next-nearest-neighbor, and the next-next-nearest-neighbor ones:
\begin{equation}
H_J=J_1 \sum_{\langle ij\rangle}S_i\cdot S_j+J_2 \sum_{\langle ij\rangle^{\prime}}S_i\cdot S_j+J_3 \sum_{\langle ij\rangle^{\prime\prime}}S_i\cdot S_j.
\end{equation}
Here $\langle ij\rangle$, $\langle ij\rangle^{\prime}$, $\langle ij\rangle^{\prime\prime}$ represent the summation over the nearest, next-nearest, and next-next-nearest neighbors, respectively. We set $J_1:J_2:J_3=-1:1.676:0.997$ with $J_1<0$, from the first principle calculation~\cite{yan}.

The bare spin susceptibility from the tight-banding part can be calculated as,
\begin{equation}
\chi_0({\bf q},\omega)=\frac{1}{N}\sum_{\bf k}\frac{f(\varepsilon_{{\bf k}+{\bf q}})-f(\varepsilon_{{\bf k}})}{\omega-(\varepsilon_{{\bf k}+{\bf q}}-\varepsilon_{{\bf k}})+i\Gamma},
\end{equation}
where $f(x)$ is the Fermi distribution function.

The correction of the spin fluctuation from the spin-spin interaction $H_J$ is included in the random phase approximation (RPA),
\begin{equation}
\chi({\bf q},\omega)=\frac{\chi_0({\bf q},\omega)}{1-J_{\bf q}\chi_0({\bf q},\omega)},
\end{equation}
where $J_{\bf q}=-J_1 (\cos k_x+\cos k_y)-2J_2 \cos k_x \cos k_y-J_3 (\cos 2k_x+\cos 2k_y)$, is the Fourier factor of the spin coupling term $H_J$.

Considering that the SC pairing is meditated by the spin fluctuation, we may write the linearized eliashberg's equation,
\begin{equation}
\lambda \Delta({\bf k})=-\sum_{\bf k^{\prime}}V({\bf k}-{\bf k^{\prime}})\frac{\tanh(\beta \varepsilon_{\bf k^{\prime}}/2)}{2\varepsilon_{\bf k^{\prime}}}\Delta({\bf k^{\prime}}),
\end{equation}
with $\beta=1/T$. We consider the spin-fluctuation as the
effective pairing potential $V({\bf q})=J^2_{\bf q}\chi({\bf q},0)$. Since we here address the pairing symmetry,  we  focus only on the zero-energy spin susceptibility, which should produce qualitatively correct results for the pairing symmetry as usual~\cite{kuro,xsye}.

\begin{figure}
\centering
  \includegraphics[width=8cm]{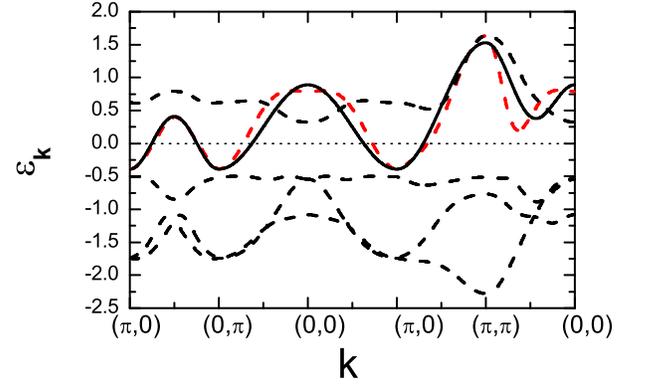}
\caption{(Color online) The band structure along different momentum cut. The solid line is the calculated band structure from our single-band model. The dashed lines are the replot of the band structure for the five-orbital model with the parameters in Ref.~\cite{yu}.}
\end{figure}

We first depict in  Fig.1 the band dispersion calculated from the five-band model for KFe$_2$Se$_2$ proposed in Ref~\cite{yu}. As is seen, there are three bands below the Fermi energy and one band above it, with the energies less than $-0.5$ eV and larger than $0.25$ eV. The SC gap is $\sim 0.01$ eV from the ARPES experiments~\cite{zhang,mou,zhao,xpw}. As a result, these bands (not crossing the FS) should contribute little to the superconductivity. It is reasonable to concentrate on the band that crosses the FS, which plays an essential role in the superconductivity.
 Considering that the whole electron filling is $6+\delta$, the three filled bands contribute six electrons per site
 and thus the electron filling for the band that crosses the Fermi energy is $\delta$. These features are crucially captured by the the present single-band model. The corresponding bare band dispersion is plotted in Fig.1, which almost coincide with the crossing Fermi energy band. This similarity suggests that  our model could produce the qualitatively same results as those obtained from the five-orbital model at low energies.
 %Although only one band is considered here, it is actually contributed by all five orbits, which are hybridized strongly.

We now study the spin fluctuation based on the above single-band model at the doping $\delta=0.4$. As mentioned above, the spin susceptibility are mainly determined from the two factors:  one is the bare spin susceptibility $\chi_0$ evaluated from the tight-banding term, the other the RPA factor $1-J_{\bf q}\chi_0$ closely related to the local spin interaction.
The behavior of the bare spin susceptibility is usually determined by the FS topology, with the maximum spin excitation occurring at near the wave vector connecting different FS sheets. For the renormalized one, the pattern of $J_{\bf q}$, representing spin interaction strength in the Fourier space, plays an important role. Obviously, a larger $J_{\bf q}$ will lead to a larger spin susceptibility (The value of $J_1$ in $J_{\bf q}$ is usually determined from the SDW instability point). Since our main conclusions do not depend on $|J_1|$, we take $J_1=-0.1$ eV for illustration.
The bare and renormalized spin susceptibility along the two-dimensional cut are plotted in Fig.2(a). The Fourier factor of the spin interaction $J_{\bf q}$ is displayed in Fig.2(b). Figs.2(c) and 2(d) are the intensity plots of the bare and renormalized spin susceptibility in the whole BZ, respectively.
As is seen, the largest spin susceptibility occurs at the wave vector $(\pi/2,\pi/2)$. This result, elaborated below to be consistent with the FS topology,
   may be checked by the future neutron scattering experiments.

\begin{figure}
\centering
  \includegraphics[width=8cm]{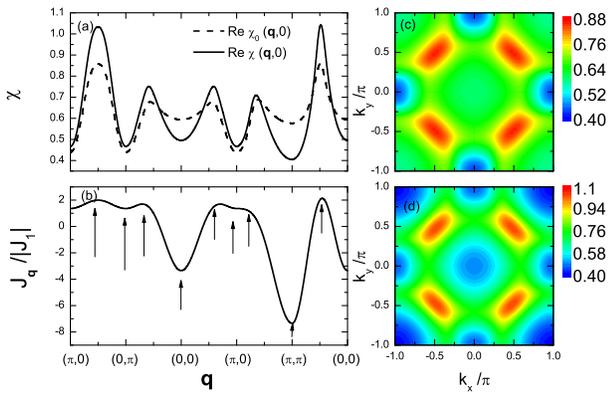}
\caption{(Color online) (a) The bare and renormalized zero energy spin susceptibility with $J_1=-0.1$ eV along different momentum cut.
(b) The Fourier factor $J_{\bf q}$ as a function of the momentum ${\bf q}$. Panels (c) and (d) are the intensity plot of the bare and renormalized spin susceptibility, respectively.   }
\end{figure}

An important feature can be drawn from the results of the spin fluctuation. Notably, there exists an implicit relationship between the bare spin susceptibility and the term $J_{\bf q}$ when comparing Figs.2(a) with 2(b), namely, $J_{\bf q}$ and $\chi_0$ have a qualitatively same momentum dependence.
As a result, the RPA correction increases the spin fluctuation, strengthens the peaks, and suppresses the dips of the bare one. Since the superconductivity is expected to be meditated by the spin fluctuation,  a stronger fluctuation means generally the stronger superconductivity.
The matching of the vertex term $J_{\bf q}$ and the bare band spin excitation provides a natural explanation for why the SC transition temperature is high for this compound. On the other hand, due to this implicit relationship,
the understandings of the appearance of SDW (AF) instability based respectively on the fermiology and the local spin interaction picture are also consistent. Actually, for the previous high-T$_c$ SC materials, a similar relationship exists, e.g., the bare spin excitation is peaked at $(\pi,\pi)$ for cuprates and consistent with the factor $J_{\bf q}=-J(\cos q_x+\cos q_y)$ when the nearest-neighbor spin interaction is considered. For iron-pnictide,   $J_{\bf q}$ in the $J_1-J_2$ model is usually written as $J_{\bf q}=-J_1(\cos q_x+\cos q_y)-2J_2 \cos q_x\cos q_y$, which has the maximum at $(\pi,0)$ or $(0,\pi)$ if  $J_2>J_1/2$, noting that $J_2>J_1/2$ is supported by the first principle calculation~\cite{yil,fma,qsi}. Meanwhile, the spin susceptibility contributed from the bare band reaches its maximum at $(\pi,0)$ and $(0,\pi)$. Thus this relationship between  $J_{\bf q}$ and the bare spin susceptibility  enhances the spin fluctuation in the framework of RPA and correspondingly the superconductivity.

\begin{figure}
\centering
  \includegraphics[width=8cm]{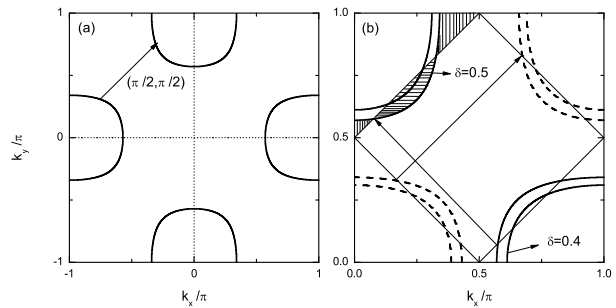}
\caption{(a) The FS in the whole BZ. (b) The replot of FSs in the upper right 1/4-BZ with doping $\delta=0.4$ and $0.5$, respectively.
The dashed lines are the $(\pi,0)$-shift of the FSs from the upper left 1/4-BZ}
\end{figure}

We now turn to address the origin of the $(\pi/2,\pi/2)$ spin fluctuation based on the fermiology picture.
The FS with doping $\delta=0.5$ is plotted in Fig.3(a). The FS sheets around $X$ and its symmetric points are clearly seen. The wave vector connecting the different FS sheets is just  ${\bf Q}=(\pi/2,\pi/2)$. To compare the present one with the cuprates and to get an intuitive understanding of the interplay among the fermiology, the spin fluctuation, and superconductivity, we
replot the FS sheets in the upper right BZ in Fig.3(b). The dashed lines are the $(\pi,0)$ shift of the FS sheets from the upper left BZ. We can define the crossing points of the FS with the lines ($k_x=k_y\pm\pi/2$) as the hot spots. Then the wave vector connecting the hot spots on different FS sheets, shown in the figure, are the wave vector ${\bf Q}$.
The distance between the different FS sheets is close to ${\bf Q}$.
For the parent compound, the area of vertical shadowed region should equal to that of horizontal shadowed region. The wave vector connecting the FS sheets should be close to $(\pm\pi/2,\pm\pi/2)$. Based on the fermiology and Eq.(3), it is rather clear that the spin susceptibility reaches its largest value at ${\bf Q}$ because $\varepsilon_{{\bf k}+{\bf Q}}-\varepsilon_{{\bf k}}$ is vanishingly small.
For the doped systems, the balance between the two kinds of shadow is broken,
and
the hot spots shift to the BZ boundary or diagonal direction.
The features of  the FS topology and doping evolution seen from Fig.3(b) are quite similar to the FS of cuprates
in the whole BZ, except that  the wave vectors connecting the hot spots are $(\pm\pi/2,\pm\pi/2)$ for this compound by noting that the corresponding vectors are $(\pm\pi,\pm\pi)$ for cuprates. In Fig.3(b),
only two FS sheets actually exist.
For cuprates four FS sheets exist. As a result, the electron filling for the parent compound is only half to that for cuprates, namely, only $0.5$ for the parent KFe$_2$Se$_2$.

Besides the fermiology, it is insightful to compare the present material with cuprates through the Heisenberg spin interaction term $H_J$.
For cuprates, only the nearest-neighbor interaction needs to be considered with $J_1>0$, leading to the AF ground state.
When the AF ground state is suppressed, the spin excitation occurs at the wave vector near AF wave vector $(\pi,\pi)$.
For the present material, the parent compound is modeled by the $J_1-J_2-J_3$ Hamiltonian~\cite{yan}. Here $J_1<0$ and $J_2$, $J_3$ can be comparable to $J_1$, which is quite different from that of cupates.
The ground state is frustrated and the combined effect from all $J_i$-term leads to the block AF state, with one block consisting of $2\times 2$ sites.
The period along $x$ or $y$ direction is twice of that of cuprates.
Thus when the spin order is suppressed,  the wave vector of the spin excitation is half of that in cuprates,
namely,  $(\pi/2,\pi/2)$.
%On the other hand, as addressed, in spite that the FS topology and spin excitation are different,
%the interplay of the superconductivity and magnetism should be similar to cuprates.

\begin{figure}
\centering
  \includegraphics[width=8cm]{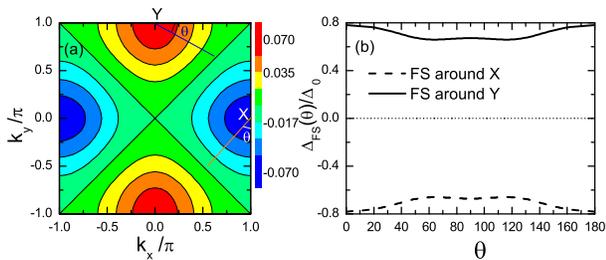}
\caption{(Color online) (a) The gap function from the self-consistent calculation. (b) The gap along the FS sheets with the dashed and solid lines are around the X and Y points, respectively, with $\theta$ denoted in panel (a).}
\end{figure}

At this stage, we look into the pairing symmetry of
the SC gap, which can be evaluated from Eq.(5), namely, the temperature $T$ is decreased until the maximum eigenvalue $\lambda=1$ is obtained, with the SC gap being the
eigenvector for the maximum eigenvalue. The intensity plot of the SC gap is displayed in Fig.4(a). Obviously, the pairing symmetry is of $d_{x^2-y^2}$-wave. We have checked numerically that this result is rather robust to the reasonable parameter change. The SC gap along the FS for the $d$-wave pairing symmetry $\Delta=\Delta_0/2(\cos k_x-\cos k_y)$ is plotted in Fig.4(b).
As is seen, the SC gap along one FS sheet is nearly isotropic with the difference less than 15\%, being  consistent with the ARPES experiments~\cite{zhang,mou,zhao,xpw}. In addition, it was reported in ARPES experiments that the gap is small or vanish for the smaller pockets around the $\Gamma$ point~\cite{zhang,mou}, being also qualitatively consistent with the $d_{x^2-y^2}$-wave. However, it was indicated
in Refs.~\cite{mou,xpw} that there exists a larger FS pockets around $\Gamma$ with the gap being nearly isotropic. This result seems to contradict with the $d$-wave symmetry. While we think that the larger FS around $\Gamma$ is actually due to the band folding effect: dominant features of the pocket and the gap may be folded from those around $M$ point. If this is the case, our results for the gap symmetry agree qualitatively with the ARPES experiments.

The origin of the $d$-wave superconductivity can be understood based on the spin fluctuation picture and FS topology. The pairing potential $V$ contributed by the spin fluctuation is largest at the wave vector ${\bf Q}=(\pm\pi/2,\pm\pi/2)$. The factor ${\tanh(\beta \varepsilon_{\bf k^{\prime}}/2)}/{2\varepsilon_{\bf k^{\prime}}}$ in Eq.(5) is positive for any $\varepsilon_{\bf k^{\prime}}$ and largest at $\varepsilon_{\bf k^{\prime}}=0$, which
means that the pairing near the FS is important.
As a result, the gap function at or near the FS should satisfy the condition $\Delta_{\bf k}=-\Delta_{{\bf k}+{\bf Q}}$. As shown in Fig.3, if ${\bf k}$
belongs to one sheet of FS, then ${\bf k}+{\bf Q}$ should be near the other neighboring sheet of FS. For the $d_{x^2-y^2}$-symmetry, as displayed in Fig.4(b), the SC gaps have the same magnitudes and different signs along the two neighboring FS sheets. The condition $\Delta_{\bf k}=-\Delta_{{\bf k}+{\bf Q}}$ is satisfied approximately. In this sense, we give an intuitive understanding of the pairing symmetry in this material.

In summary, we have proposed a single-band model to describe the K$_x$Fe$_2$Se$_2$ material. The band structure has been found to be qualitative consistent with the five-orbital model at low energies. The spin excitation with the wave vector $(\pm\pi/2,\pm\pi/2)$ has been revealed. Our theoretical results can well be understood based on the dierct fermiology analysis. % and consistent with the $J_1-J_2-J_3$ model based on the first principle calculation.
 In particular, we have presented a sound and coherent picture for the interplay of the SDW and superconductivity in this compound and suggested that the essential physics is somehow similar to that in other high-T$_c$ SC materials. Finally, the rather robust $d_{x^2-y^2}$ pairing symmetry of superconductivity has been revealed.
%which can be understood based on the fermiology theory.

This work was supported by the NSFC under the Grant
No. 11004105, the RGC of Hong Kong under the No.
HKU7055/09P and a CRF of Hong Kong.


\begin{thebibliography}{99}
\bibitem{kam} Y. Kamihara {\it et al.}, J. Am. Chem. Soc. {\bf 130}, 3296 (2008).
\bibitem{sad} M. V. Sadovskii, Uspekhi Fiz. Nauk {\bf 178}, 1243 (2008)
\bibitem{ish} K. Ishida, Y. Nakai, H. Hosono. J.Phys. Soc. Jpn. {\bf 78},
062001 (2009).
\bibitem{jiang} Q. Han, Y. Chen, and Z. D. Wang, Europhy. Lett. {\bf 82}, 37007 (2008); H. M. Jiang, J. X. Li, and Z. D. Wang, Phys. Rev. B {\bf 80}, 134505
(2009).
\bibitem{zho} Tao Zhou, Degang Zhang, and C. S. Ting, Phys. Rev. B {\bf 81},
052506 (2010).
\bibitem{yil} T. Yildirim, Phys. Rev. Lett. {\bf 101}, 057010 (2008).
\bibitem{fma} F. Ma, Z.-Y. Lu, and T. Xiang, Phys. Rev. B {\bf 78}, 224517
(2008).
\bibitem{qsi} Q. Si and E. Abrahams, Phys. Rev. Lett. {\bf 101}, 076401
(2008).
\bibitem{mazin} I. I. Mazin, D. J. Singh, M. D. Johannes, and M. H. Du, Phys.
Rev. Lett. {\bf 101}, 057003 (2008).
\bibitem{yao} Z.-J. Yao, J. X. Li, and Z. D. Wang, New J. Phys. {\bf 11},
025009 (2009).
\bibitem{guo} J. Guo {\it et al.}, Phys. Rev. B {\bf 82}, 180520(R) (2010).
\bibitem{mazio} A. Krzton-Maziopa {\it et al.}, J. Phys.: Condens. Matter {\bf 23}, 052203 (2011).
\bibitem{fang} Minghu Fang {\it et al.}, Europhy. Lett.  {\bf 94}, 27009 (2011).
\bibitem{zhang} Y. Zhang {\it et al.}, Nature Materials {\bf 10}, 273 (2011).
\bibitem{mou} D. Mou {\it et al.}, Phys. Rev. Lett. {\bf 106}, 107001 (2011).
\bibitem{zhao} L. Zhao {\it et al.}, Phys. Rev. B {\bf 83}, 140508(R) (2011).
\bibitem{xpw} X.-P. Wang {\it et al.}, Europhy. Lett. {\bf 93}, 57001 (2011).
\bibitem{qian} T. Qian {\it et al.}, Phys. Rev. Lett. {\bf 106}, 187001 (2011).
\bibitem{she} I.R. Shein and A.L. Ivanovskii, arXiv:1012.5164.
\bibitem{yan} Xun-Wang Yan {\it et al.}, arxiv:1012.5536.
\bibitem{cao} Chao Cao and Jianhui Dai, Chi. Phys. Lett., {\bf 28}, 057402 (2011).
\bibitem{sek} Y. Sekiba {\it et al.}, New J. Phys. 11, 025020 (2009).
\bibitem{yu} Rong Yu {\it et al.}, arXiv:1103.3259.
\bibitem{wang} F. Wang {\it et al.}, Europhy. Lett. {\bf 93}, 57003 (2011).
\bibitem{mai} T. A. Maier {\it et al.}, Phys. Rev. B {\bf 83}, 100515(R) (2011).
\bibitem{kot} H. Kotegawa {\it et al.}, J. Phys. Soc. Jpn. {\bf 80}, 043708 (2011).
 \bibitem{sai} T. Saito, S. Onari, and H. Kontani, Phys. Rev. B {\bf 83}, 140512(R) (2011).
 \bibitem{yzhou} Yi Zhou {\it et al.}, Europhy. Lett. {\bf 95}, 17003 (2011).

\bibitem{kur} Kazuhiko Kuroki {\it et al.}, Phys. Rev. Lett. {\bf 101}, 087004 (2008).
\bibitem{rag} S. Raghu {\it et al.}, Phys. Rev. B {\bf 77}, 220503(R) (2008).
\bibitem{ryu} Rong Yu, Jian-Xin Zhu, and Qimiao Si, Phys. rev. Lett. {\bf 106}, 186401 (2011).
\bibitem{note} In this minimum model, we have neglected a possible vacancy order, which may occur as the Fe content is reduced. It has been believed widely that the vacancy may merely produce some shadow bands with low spectral weights, which are not relevant to the superconductivity.
\bibitem{kuro} K. Kuroki, Y. Tanaka, and R. Arita, Phys. Rev. B {\bf 71}, 024506
(2005).
\bibitem{xsye} X. S. Ye and J. X. Li, Phys. Rev. B {\bf 76}, 174503 (2007).



\end{thebibliography}
\end{document}